\documentclass[useAMS,usenatbib,usegraphicx]{mn2e}
\usepackage{times}
\usepackage{amssymb}

\title[]{A Holistic Abundance Analysis to r-rich Stars}
\author[Jiang Zhang , Wenyuan Cui  and Bo Zhang]{Jiang Zhang$^{1,2,3}$,
Wenyuan Cui$^{1,2}$ and Bo Zhang$^{1,2}$\thanks{Corresponding
author.
Email-address: zhangbo@mail.hebtu.edu.cn (Bo Zhang)}\\
$^{1}$Department of Physics, Hebei Normal University, 113 Yuhua
Dong Road, Shijiazhuang 050016, China\\
$^{2}$ Hebei Advanced Thin Films Laboratory, Shijiazhuang 050016\\
$^{3}$School of Mathematics and Science, Shijiazhuang University of
Economics, Shijiazhuang 050016, China}
\begin{document}

\date{Received...; in original form...}

\pagerange{\pageref{firstpage}--\pageref{lastpage}} \pubyear{2009}

\maketitle

\label{firstpage}

\begin{abstract}
The chemical abundances of the metal-poor star are an excellent test
bed to set new constraints on models of neutron-capture processes at
low metallicity. Some r-rich metal-poor stars, such as HD 221170,
show overabundance of the heavier neutron-capture elements and
excesses of lighter neutron-capture elements. The study for these
r-rich stars could make us get a better understanding of weak r- and
main r-process nucleosynthesis at low metallicity. Based on the
conclusions of the observation of metal-poor stars and
neutron-capture element nucleosynthesis theory, we set up a model to
determine the relative contributions from weak r- and main r-process
to the heavy element abundances in metal-poor stars. Using this
model, we find that the abundance patterns of light elements for
most sample stars are close to the pattern of the weak r-process
star, and heavier neutron-capture elements is very similar to main
r-process star, while the lighter neutron-capture elements can be
fitted by mixing of weak r- and main r-process material. The
production of the weak r-process elements appears to be associated
with the light elements and the production of main r-process
elements are almost decoupled from that of the light elements. We
compare our results with the observed data at low metallicities,
showing that the predicted trends are in good agreement with the
observed trends, at least for the metallicity range [Fe/H]$<$$-2.1$.
For most of sample stars, the abundance pattern of both
neutron-capture elements and light elements could be best explained
by a star formed in a molecular cloud that had been polluted by both
weak r- and main r-process material.

\end{abstract}


\section{Introduction}

The elements heavier than the iron peak are made through neutron
capture via two principal processes: the r-process and the
s-process (Burbidge et al. 1957). Observational evidence and
theoretical studies have identified the main s-process site in
low- to intermediate-mass ($\approx1.3-8M_{\odot}$) stars in the
asymptotic giant branch (AGB) (Busso et al. 1999). The r-process
is usually associated with the explosive environment of Type II
supernovae (SNeII), although this astrophysical site has not been
fully confirmed yet (Arnould et al. 2007, Sneden et al. 2008). The
abundances of neutron-capture elements and other light elements in
metal-poor halo stars are now providing important clues to the
chemical evolution and early nucleosythesis history of the Galaxy.

The process producing heavier r-process elements ($Z\geq56$, i.e.,
heavier than Ba) is referred to as main r-process (e.g., Truran et
al. 2002; Wanajo \& Ishimaru 2006). The observations in the
ultra-metal-poor halo star CS 22892-052 (Sneden et al. 2003) and
CS 31082-001(Hill et al. 2002; Honda et al. 2004; Plez et al.
2004; Barklem et al. 2005; Spite et al. 2005) revealed that the
abundances of the heavier stable neutron-capture elements are in
remarkable agreement with the scaled solar system r-process
pattern, called as ``main r-process stars" (Sneden et al. 2003).
More recent work , utilizing updated experimental atomic data to
determine more accurate abundances, has confirmed this agreement
(Sneden et al. 2009). Since they are thought to exhibit the
abundance pattern produced by a single or at most a few r-process
events in the early Galaxy, the stability of the observed
abundance pattern of neutron-capture elements and the good
agreement with the solar system heavier r-process contribution
imply that main r-process events generate a universal abundance
distribution. Because the strong enhancements of heavier r-process
elements relative to iron and other light elements, it is believed
that the production of main r-process elements is fully decoupled
from that of Fe and all other light elements between N and Ge
(Qian \& Wasserburg 2007). However, these objects show some
deficiencies in lighter neutron-capture elements ($37\leq Z
\leq47$, i.e., from Rb to Ag) compared to the scaled solar
r-process curve (e.g., Sneden et al. 2000; Hill et al. 2002). This
means that the r-process abundance pattern in solar-system
material is not fully explained by the component that makes the
pattern found in such main r-process-enhanced stars, but another
neutron-capture process referred sometimes as ``LEPP" (lighter
element primary process) or ``weak r-process" is required
(Travaglio et al. 2004; Wanajo \& Ishimaru 2006). In addition to
the neutron-capture elements abundance distribution, observations
of other elemental abundances in unevolved metal-poor halo stars
can also provide important clues about nucleosynthesis events in
the early Galaxy. These stars are old and preserve in their
photospheres the abundance composition at the location and time of
their formation. Thus, the abundances of the light elements in
main r-process stars should reveal the composition of interstellar
medium (ISM) at the location and time of their formation.

The moderate enhancements of Eu and other r-process elements
relative to iron have been observed in normal metal-poor stars,
which means that the abundance of Eu ([Eu/Fe]$\approx$0.3) is
enhanced moderately in ISM (Fields et al 2002). However, some very
metal-poor stars, such as HD 122563, have excesses of lighter
neutron-capture elements Sr, Y, and Zr, while their heavier ones
(e.g., Ba, Eu) are very deficient ([Eu/Fe]$\approx-0.5$)
(McWilliam 1998; Johnson 2002; Aoki et al. 2005; Honda et al.
2004, 2006, 2007; Spite et al. 2005; Andrievsky et al. 2007). Such
objects possibly record the abundance patterns produced by another
component and are called as ``weak r-process stars" (Izutani et
al. 2009). In HD 122563, the abundances of neutron-capture
elements continuously decrease with the increase of atomic number
and the lighter neutron-capture elements with intermediate mass
elements show moderate enhancements with respect to heavier ones.
This is the first determination of the overall abundance pattern
that could represent the yields of the weak r-process. Recently,
Honda et al. (2007) showed another example of the weak r-process
stars HD 88609, and concluded that the abundance pattern of
neutron-capture elements in HD 88609 is quite similar to HD
122563. However, the similarity of the abundance pattern found in
the two objects can not be interpreted as uniqueness of the
pattern produced by the weak r-process (or LEPP), because the two
objects were selected to have similarly high Sr/Eu abundance
ratios. In the cloud polluted by weak r-process,
[Eu/Fe]$\approx-0.5$ means that the production of Fe group and
other light elements is decoupled from heavier r-process elements,
while [Sr/Fe]$\approx0$ implies that the weak r-process elements
are produced in conjunction with the Fe and light elements. Thus,
the abundances of both lighter neutron-capture and light elements
in weak r-process stars should reveal the composition of the cloud
polluted by weak r-process event. In this case, the holistic
abundance pattern including both the light and weak r-process
elements should be considered.

Based on observations of metal-poor stars having different Sr/Eu
ratios, Montes et al.(2007) concluded that the weak r-process
produces a uniform and unique abundance pattern of neutron-capture
elements. They found that the uncertainties on the weak r-process
pattern obtained are smaller for stars with significant weak
r-process contribution, such as HD 122563 and HD 88609, but larger
for higher [Eu/Fe]. When [Eu/Fe] reaches 0.8, the uncertainties will
reach about 1.0 dex (see Figure 3 in Montes et al. 2007). Therefore
to obtain more information about weak r-process, one needs to
investigate the stars with the moderate overabundance in [Eu/Fe]
where both the weak r- and main r-process do not prominently
dominate the composition. Especially, the moderate r-process
enhanced metal-poor stars with [Eu/Fe]$\gtrsim$0.5 should be very
important for this topic.

Recently, Ivans et al. (2006) analyzed the spectra of the metal-poor
star HD 221170 ([Eu/Fe]=0.8). Surprisingly, in contrast to the
abundance patterns of other main r-process stars, the abundances of
HD 221170 do not show the pronounced underproduction of lighter
r-process elements as seen in other main r-process enhanced stars,
and could be well fit by solar r-process abundance pattern. The
[Eu/Fe] of HD 221170 is 0.8, which just means that the weak
r-process do not prominently dominate the composition. There have
been many theoretical studies of r-process nucleosynthesis.
Unfortunately, the precise r-process nucleosynthesis sites or what
the sites for these various mass ranges of neutron-capture elements
observed in the halo stars are still unknown. As the abundances of
the metal-poor halo stars can be applied as a probe of the
conditions of r-process nucleosynthesis in the early Galaxy,
clearly, the holistic analysis of both neutron-capture and other
light elemental abundances in HD 221170 and other metal-poor stars
is very important for a good understanding of the neutron-capture
nucleosynthesis in the early Galaxy. These reasons motivated us to
investigate the holistic element abundance patterns in the
metal-poor r-process-rich (hereafter r-rich) stars, in which light
elements, lighter and heavier neutron-capture elements are observed.

In this paper, we study elemental abundances of 14 metal-poor stars,
in which more than about 20 elements have been observed. Firstly, we
extend the abundance pattern of main r-process and weak r-process to
light elements. Then, we investigate the characteristics of the
nucleosynthesis that produces the abundance ratios of these stars
using the weak r- and main r-process parametric model. In Sect.\ 2
we extracted the abundance clues about weak r- and main r-process
from the metal-poor star HD 2221170. The parametric model of
metal-poor stars is described in Sect.\ 3. The calculated results
are presented in Sect.\ 4 which we also discuss the characteristics
of the weak r- and main r-process nucleosynthesis. Conclusions are
given in Sect.\ 5.

\section{Abundance Clues}

Recent observations and the analyses imply that the abundance
pattern of an extremely metal-deficient star with [Fe/H]$\leq-2.5$
may retain information of a preceding single supernova (SN) event
or at most a few SNe (Mcwilliam et al. 1995; Ryan et al. 1996). It
is important to simultaneously analyse the observed light and
heavy elements to study the physical conditions that could
reproduce the observed abundance pattern found in metal-poor
stars. As an example, we investigate the observed elemental
abundance pattern of metal-poor HD 221170 (Ivans et al. 2006).

\begin{figure*}
 \centering
 \includegraphics[width=1.0
 \textwidth,height=0.5\textheight]{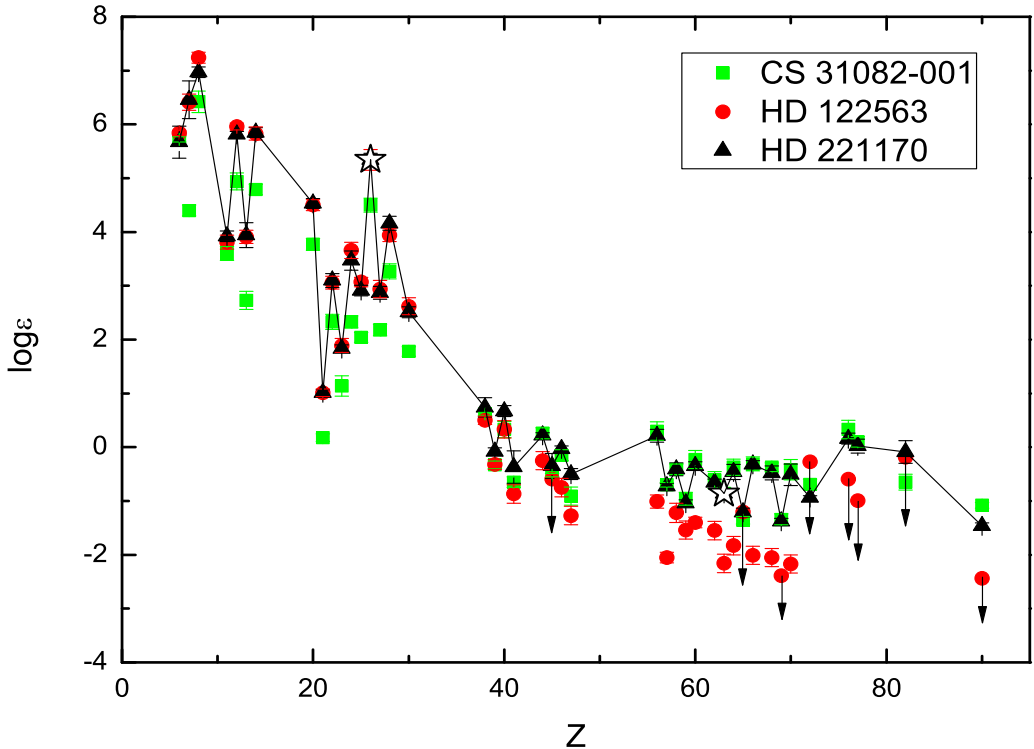}
\caption{Abundance pattern of HD 221170, scaled abundance pattern
obtained by main r-process star CS 31082-001, and scaled abundance
pattern obtained by weak r star HD 122563. The CS 31082-001 was
normalized to the Eu abundance while HD 122563 was normalized to the
Fe abundance. The scaled Fe and Eu have been denoted specifically by
open stars.}
\end{figure*}

Figure 1 shows the abundance comparisons of the abundance pattern
of HD 221170 with that of the weak r-process star HD 122563 and
the main r-process star CS 31082-001 in the logarithmic scale. The
abundances of CS 31082-001 are normalized to Eu, which is
predominantly an r-process element. For the two stars, HD 221170
and CS 31082-001, shown in this figure, have almost identical
abundance pattern from Ba to Ir, but the light elements, such as
N, O, Mg and Fe group, differ greater than 0.5 dex, which
indicates that the origin of light elements and main r-process is
seemly different. Because the excess of light elements for HD
221170, could not be fitted simultaneously by the abundances of
main r-process star, we can conclude that the production of light
elements should not accompany with the production of the heavier
r-process elements, which is similar to the abundance
characteristic of main r-process stars (Qian \& Wasserburg 2007).
On the other hand, we take HD 122563 as a weak r-process star
(Honda et al. 2006), and plot its observed element abundances
normalized to Fe, because Fe group elements are produced in
conjunction with the weak r-process. Our conclusion here is that
the abundance pattern of light elements in HD 221170 is quite
similar to that of HD 122563, although a small difference is
suggested. In summary, we have found that the abundance pattern of
HD 221170 can be fitted partly by abundance curves from the weak
r-process stars and the main r-process stars. Obviously, a precise
calculation is needed.

\section{Parametric Model of Metal-Poor Stars}

We note that considering the larger overabundance of Eu to Fe for
main r-process star and deficient of Eu to Fe for weak r-process
star, the origins of the light elements in two stars are clearly
different. For main r-process stars, the light elements mainly
come from the cloud in which the star formed, while for weak
r-process stars, the light elements mainly come from the
production in conjunction with the weak r-process. Since the main
r-process are not sufficient to explain the observed abundances in
HD 221170 and some other metal-poor stars, another component that
yields both lighter neutron-capture elements and light elements is
required. We assume that all of the r-process abundances come from
two r-processes, each with a different set of abundance
signatures. It is interesting to adopt the two r-processes
components parametric model to study the relative contributions
from the weak r- and main r- processes that could reproduce the
observed abundance pattern found in r-rich stars. For this
purpose, we propose that the $ith$ element's abundance in this
star can be calculated as follows:
\begin{equation}
N_{i}=C_{w}^{'}N_{i,\ rw}+C_{m}^{'}N_{i,\ rm},
\end{equation}
where $N_{i,\ rw}$ is the abundance of the $ith$ element produced by
the weak r-process and $N_{i,\ rm}$ is the abundance of the $ith$
element produced by the main r-process, and $C_{w}^{'}$ and
$C_{m}^{'}$ are the component coefficients that correspond to
contributions from the weak r-process and the main r-process,
respectively.

The ultra-metal-poor stars CS 22892-052 and CS 31082-001 merit
special attention, because these two stars have extremely large over
abundances of neutron-capture elements relative to iron and very low
metallicity with [Fe/H]$\sim-3$. Many studies (Cowan et al. 1999;
Sneden et al. 1996, 1998, 2000) have suggested that the abundance
patterns of the heavier stable neutron-capture elements in these
stars are consistent with the solar system r-process abundance
distribution. However, this concordance breaks down for the lighter
neutron-capture elements (Sneden et al. 2000). These two stars could
have abundances that reflect results of the pure main r-process
nucleosynthesis of a single SN, so the adopted abundances of nuclei
$N_{i,\ rm}$ in equation (1) are taken from the average abundances
of CS 22892-052 and CS 31082-001, which is normalized to the Eu
abundance of CS 22892-052. Metal-poor stars with very low Eu
abundance play an essential role in constraining the weak r-process
as they have the smallest contribution from the main r-process.
According to Montes et al. (2007), we can obtain the weak r-process
abundance pattern using average abundance of HD 122563 and HD 88609
to subtract the main r-process abundance pattern, normalized to Eu.
This means that all Eu is made in the main r-process.

From consideration of the observed abundances in metal-poor stars,
it is proposed that the sources for main r-process also produce
some lighter r-nuclei (e.g., Sr, Y, and Zr) and the production of
main r-process nuclei is not related to the production of Fe group
elements and other elements with lower atomic numbers: O, Na, Mg,
Al, Si, Ca, Sc, and Ti. However, the production of the weak
r-process elements appears to be associated with that of Fe group
elements and other light elements (Qian \& Wasserburg 2007). Thus,
we also attempt to investigate the abundance pattern of light
elements in r-rich stars utilizing equation (1). Based on the
abundances of two weak r-process stars and two main r-process
stars, we extended the abundances $N_{i,rm}$ and $N_{i,rw}$ to
light elements, except C and N, because these two elements may
have another origin in CS 22892-052 (Masseron et al. 2010). Our
goal is to find the parameters which best characterize the
observed data. The reduced $\chi^2$ is defined:
\begin{equation}
 \chi^2 = \sum \frac{(N_{obs}-N_{cal})^2}{(\bigtriangleup N_{obs})^2(K-K_{free})}, \nonumber \\
\end{equation}
where$\Delta N_{obs}$ is the uncertainty on the observed
abundance, $K$ is the number of elements applied in the fit, and
$K_{free}$ is the number of free parameters varied in the fit.
Based on equation (1), we carry out the calculation including the
contributions of the weak r- and main r- processes to fit the
abundance profile observed in HD 221170 and other metal-poor
stars, in order to look for the minimum $\chi^2$ in the two
parameter space formed by $C_{w}^{'}$, and $C_{m}^{'}$.

\section{Results and Discussion}

\begin{table*} 
 \centering
  \caption{Observed Abundance Ratios and the Derived Parameters for the sample Stars}

  \begin{tabular} {lcccccccccclrc}
  \hline
 stars &    [Fe/H] &    [Eu/Fe] &   [Sr/Fe] &   [Sr/Eu] &   $C_w^{'}$ &    $C_m^{'}$ &    $ C_{w}$ &  $ C_{m}$ &   $\chi^2$ \\
 \hline

HD 221170 & -2.18 & 0.80 &  0.02 &  -0.78 & 3.755 & 0.612 & 8.624 & 6.147 & 1.298 \\
HE 1219-0312 &  -2.96 & 1.38 &  0.35 &  -1.03 & 0.086 & 0.512 & 1.190 & 30.985 &    0.755 \\
CS 31082-001 &  -2.91 & 1.63 &  0.65 &  -0.98 & 0.009 & 0.953 & 0.111 & 51.402 &    0.434 \\
CS 29497-004 &  -2.63 & 1.68 &  0.55 &  -1.13 & 0.000 & 2.364 & 0.000 & 66.917 &    1.130 \\
CS 29491-069 &  -2.51 & 0.96 &  0.15 &  -0.81 & 1.210 & 0.651 & 5.942 & 13.979 &    1.179 \\
HD 115444 & -2.98 & 0.85 &  0.32 &  -0.53 & 0.676 & 0.126 & 9.796 & 7.985 & 1.539 \\
BD +17$^{\circ}$3248 &   -2.08 & 0.91 &  0.29 &  -0.62 & 4.315 & 1.152 & 7.872 & 9.191 & 0.903 \\
CS 22892-052 &  -3.10 & 1.64 &  0.58 &  -1.06 & 0.000 & 0.589 & 0.000 & 49.204 &    0.725 \\
HD 6268 &   -2.63 & 0.52 &  0.07 &  -0.45 & 1.438 & 0.147 & 9.308 & 4.161 & 0.933 \\
HD 122563 & -2.77 & -0.52 & -0.27 & 0.25 &  1.166 & 0.006 & 10.419 &    0.234 & 0.657 \\
HD 88609 &  -3.07 & -0.33 & -0.05 & 0.28 &  0.712 & 0.004 & 12.694 &    0.312 & 1.125 \\
HE 2224+0143 &  -2.58 & 1.05 &  0.23 &  -0.82 & 1.027 & 0.671 & 5.925 & 16.928 &    0.455 \\
HE 2252-4225 &  -2.82 & 0.99 &  0.06 &  -0.93 & 0.728 & 0.317 & 7.299 & 13.898 &    0.739 \\
HE 2327-5642 &  -2.93 & 1.22 &  0.31 &  -0.91 & 0.189 & 0.384 & 2.441 & 21.688 &    0.353 \\

\hline
\end{tabular}
\end{table*}

\begin{table*} 
 \centering
  \caption{The weak r- and main r-process elemental abundances of HD 221170}

  \begin{tabular} {lcccccccccclrc}
  \hline
Element&Z&$N_{total}$&$N_{i,rw}^{star}$&$N_{i,rm}^{star}$&$f_{r,weak}$&$f_{r,main}$\\
&  & &or&or &or& or   \\
& & &associated r-weak&associated r-main&associated r-weak& associated r-main  \\
 \hline

O   & 8   &  4.43E+05 &  3.72E+05 &  7.15E+04 &  0.839 & 0.161 \\
Na  & 11  &  2.29E+02 &  1.68E+02 &  6.07E+01 &  0.735 & 0.265 \\
Mg  & 12  &  2.01E+04 &  1.82E+04 &  1.87E+03 &  0.907 & 0.093 \\
Al  & 13  &  1.54E+02 &  1.39E+02 &  1.52E+01 &  0.901 & 0.099 \\
Si  & 14  &  1.87E+04 &  1.71E+04 &  1.65E+03 &  0.912 & 0.088 \\
Ca  & 20  &  8.60E+02 &  7.40E+02 &  1.20E+02 &  0.861 & 0.139 \\
Sc  & 21  &  2.64E-01 &  2.33E-01 &  3.08E-02 &  0.883 & 0.117 \\
Ti  & 22  &  3.11E+01 &  2.69E+01 &  4.26E+00 &  0.863 & 0.137 \\
V   & 23  &  2.80E+00 &  2.58E+00 &  2.22E-01 &  0.921 & 0.079 \\
Cr  & 24  &  1.04E+02 &  9.88E+01 &  5.58E+00 &  0.947 & 0.053 \\
Mn  & 25  &  2.34E+01 &  2.13E+01 &  2.12E+00 &  0.909 & 0.091 \\
Fe  & 26  &  5.63E+03 &  4.90E+03 &  7.28E+02 &  0.871 & 0.129 \\
Co  & 27  &  1.89E+01 &  1.56E+01 &  3.25E+00 &  0.828 & 0.172 \\
Ni  & 28  &  1.92E+02 &  1.54E+02 &  3.87E+01 &  0.799 & 0.201 \\
Cu  & 29  &  5.03E-01 &  4.18E-01 &  8.45E-02 &  0.832 & 0.168 \\
Zn  & 30  &  1.27E+01 &  1.14E+01 &  1.25E+00 &  0.901 & 0.099 \\
 Sr & 38  &  1.72E-01 &  8.12E-02 &  9.05E-02 &  0.473 & 0.527 \\
 Y  & 39  &  2.41E-02 &  1.32E-02 &  1.09E-02 &  0.549 & 0.451 \\
 Zr     & 40  &  1.22E-01 &  6.65E-02 &  5.51E-02 &  0.547 & 0.453 \\
 Nb     & 41  &  7.75E-03 &  2.89E-03 &  4.86E-03 &  0.373 & 0.627 \\
Mo  & 42  &  2.57E-02 &  1.38E-02 &  1.19E-02 &  0.536 & 0.464 \\
 Ru     & 44  &  5.24E-02 &  1.42E-02 &  3.82E-02 &  0.270 & 0.730 \\
 Rh     & 45  &  $<$1.46E-02 &  $<$7.02E-03 &  7.58E-03 &  $<$0.481 & $>$0.519 \\
 Pd     & 46  &  2.05E-02 &  4.99E-03 &  1.56E-02 &  0.243 & 0.757 \\
 Ag     & 47  &  4.54E-03 &  1.20E-03 &  3.34E-03 &  0.264 & 0.736 \\
 Ba     & 56  &  3.84E-02 &  5.34E-04 &  3.79E-02 &  0.014 & 0.986 \\
 La     & 57  &  4.31E-03 &  2.87E-05 &  4.28E-03 &  0.007 & 0.993 \\
 Ce     & 58 &  9.91E-03 &  8.52E-04 &  9.06E-03 &  0.086 & 0.914 \\
 Pr     & 59  &  3.14E-03 &  7.00E-04 &  2.44E-03 &  0.223 & 0.777 \\
 Nd     & 60  &  1.33E-02 &  3.76E-04 &  1.29E-02 &  0.028 & 0.972 \\
 Sm     & 62  &  6.71E-03 &  3.02E-04 &  6.41E-03 &  0.045 & 0.955 \\
 Eu     & 63  &  3.36E-03 &  0.00E+00 &  3.36E-03 &  0.000 & 1.000 \\
 Gd     & 64  &  1.04E-02 &  0.00E+00 &  1.04E-02 &  0.000 & 1.000 \\
Tb  & 65 &  1.60E-03 &  0.00E+00 &  1.60E-03 &  0.000 & 1.000 \\
 Dy     & 66  &  1.43E-02 &  0.00E+00 &  1.43E-02 &  0.000 & 1.000 \\
Ho  & 67  &  3.60E-03 &  0.00E+00 &  3.60E-03 &  0.000 & 1.000 \\
 Er     & 68  &  9.82E-03 &  0.00E+00 &  9.82E-03 &  0.000 & 1.000 \\
 Tm     & 69  &  1.15E-03 &  0.00E+00 &  1.15E-03 &  0.000 & 1.000 \\
Yb  & 70 &  7.53E-03 &  0.00E+00 &  7.53E-03 &  0.000 & 1.000 \\
Lu  & 71 &   2.03E-03 &  0.00E+00 &  2.03E-03 &  0.000 & 1.000 \\
Hf  & 72  &  3.66E-03 &  0.00E+00 &  3.66E-03 &  0.000 & 1.000 \\
Os  & 76  &  4.10E-02 &  0.00E+00 &  4.10E-02 &  0.000 & 1.000 \\
 Ir     & 77  &  2.90E-02 &  0.00E+00 &  2.90E-02 &  0.000 & 1.000 \\
Pt  & 78   &   4.75E-02 &  0.00E+00 &  4.75E-02 &  0.000 & 1.000 \\
Au  & 79 &    3.77E-03 &  0.00E+00 &  3.77E-03 &  0.000 & 1.000 \\
Pb  & 82  &  2.80E-02 &  0.00E+00 &  2.80E-02 &  0.000 & 1.000 \\
 Th     & 90  &  1.33E-03 &  0.00E+00 &  1.33E-03 &  0.000 & 1.000 \\
U   & 92 &  2.12E-04 &  0.00E+00 &  2.12E-04 &  0.000 & 1.000 \\

\hline

\end{tabular}
\\ Note---log$\varepsilon$(El)=logN(El)+1.54
\end{table*}

\begin{figure*}
 \centering
 \includegraphics[width=1.0
 \textwidth,height=0.9\textheight]{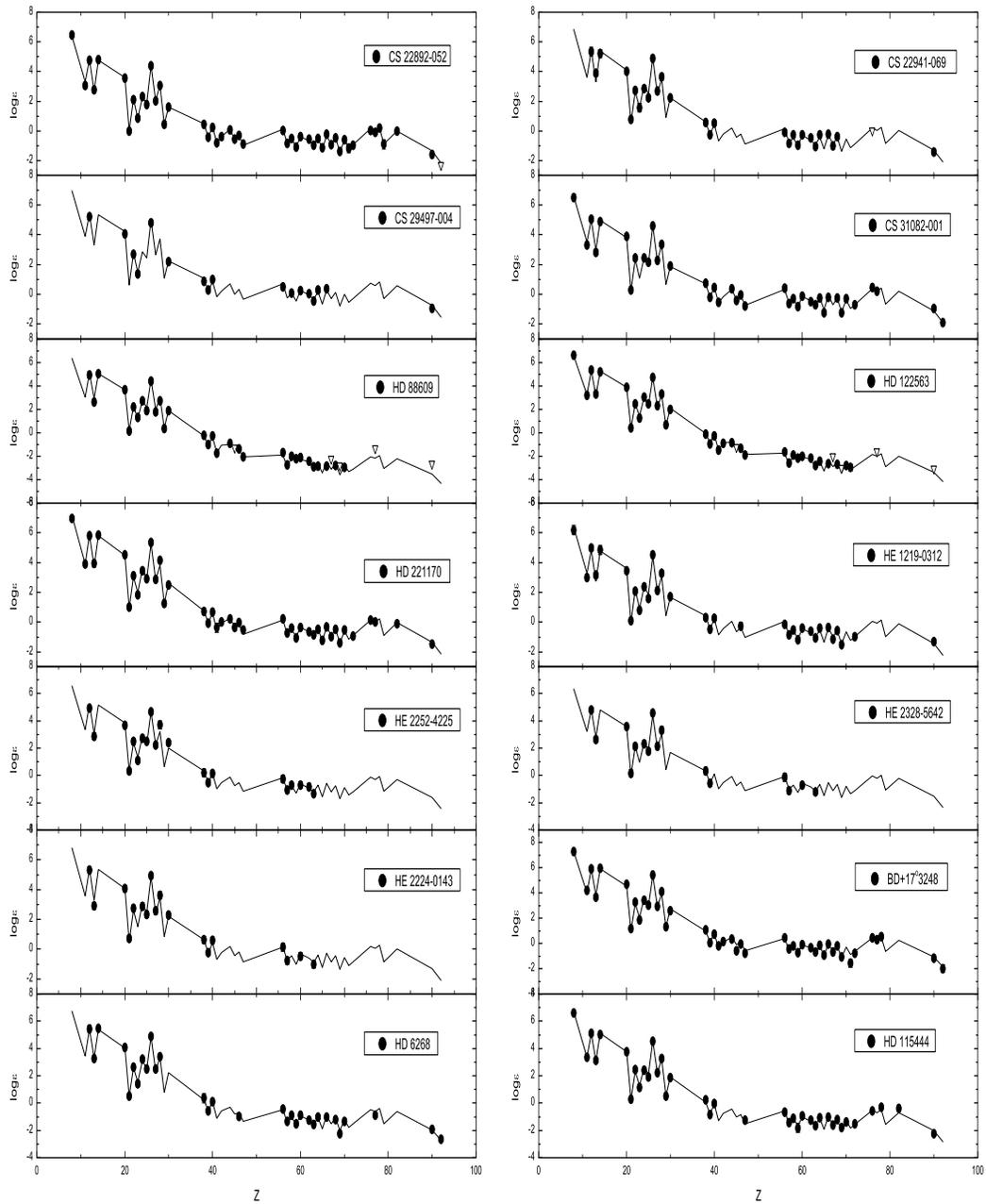}
\caption{Comparison of the observed elemental abundances for 14
sample stars with the elemental abundances calculated by this model.
The full circles with appropriate error bars denote the observed
element abundances and the solid lines are the predicted elemental
abundance curves. Upper limits are indicated by downward-facing open
triangles.}
\end{figure*}

\begin{figure*}
 \centering
 \includegraphics[width=0.88
 \textwidth,height=0.5\textheight]{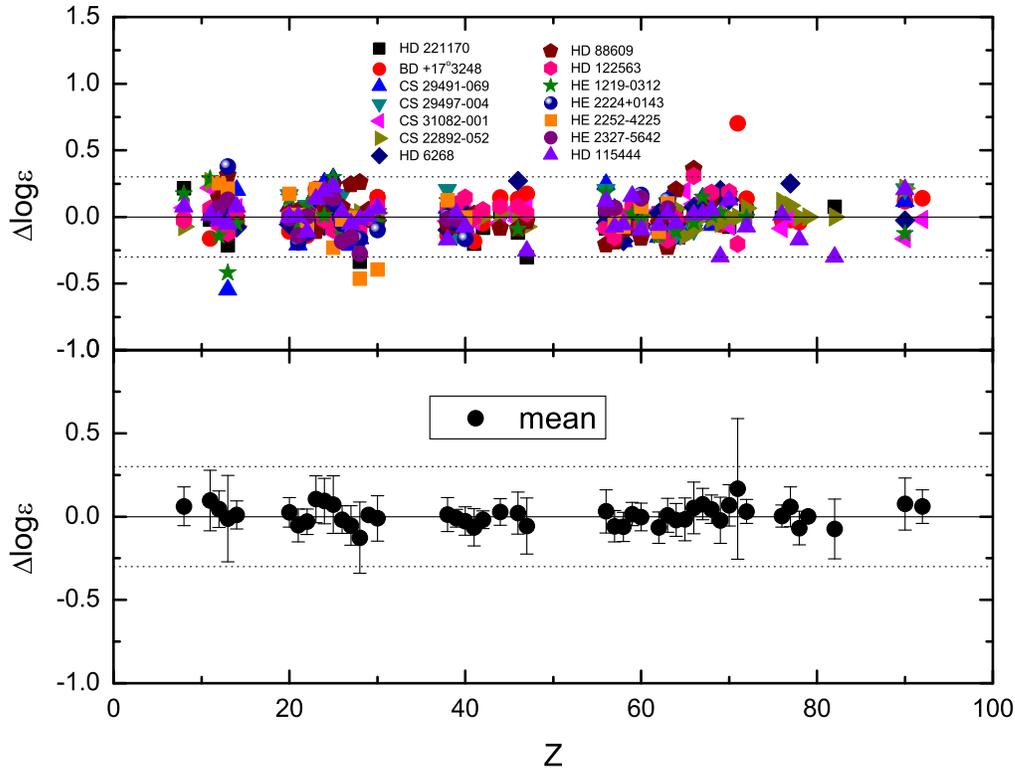}
\caption{In the top panel, difference plot showing the individual
elemental abundance offsets for each of the 14 stars with respect to
the calculated values. Zero offset is indicated by the dashed
horizontal line. The bottom panel displays average stellar abundance
offsets. }
\end{figure*}

\begin{figure*}
 \centering
 \includegraphics[width=0.8
 \textwidth,height=0.5\textheight]{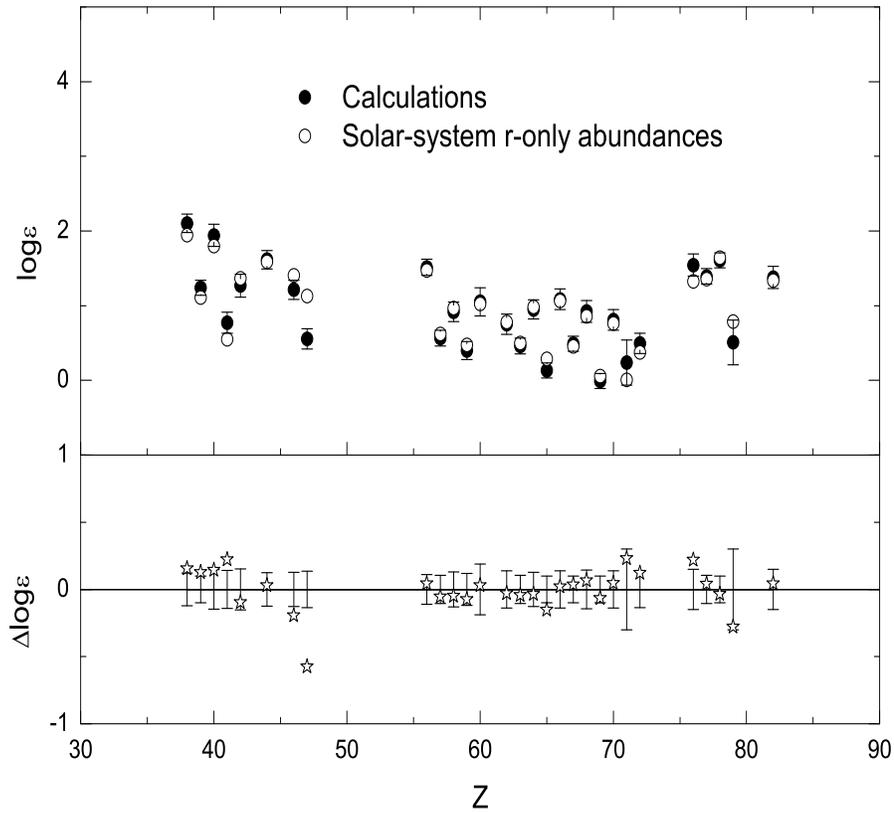}
\caption{Comparison of the calculated $log\epsilon(X)$ abundances
for $Z\geq38$ and the solar r-process predictions from Arlandini et
al.(1999). The bottom panel displays the difference defined as
$\Delta log\epsilon(X)\equiv
log\varepsilon(X)_{calc}-log\varepsilon(X)_{obs}$, where the error
bars are the average observed error of four typical stars (CS
31082-001, CS 22892-052, HD 122563 and HD 88609).}
\end{figure*}

\begin{figure*}
 \centering
 \includegraphics[width=0.88
 \textwidth,height=0.5\textheight]{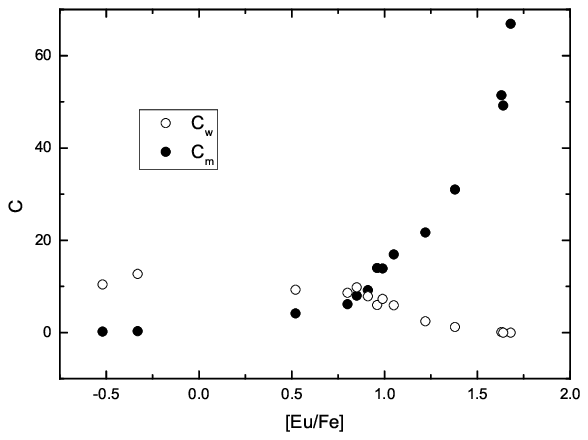}
\caption{Component coefficients of weak r- and main r-process
relative to the enrichment of Eu. }
\end{figure*}

\subsection{Calculated Results}

Using the observed data in 14 sample stars including CS 22892-052,
CS 31082-001, HD 122563 and HD 88609 (Westin et al. 2000; Cowan et
al. 2002; Hill et al. 2002; Sneden et al. 2003;  Honda et al. 2004,
2006, 2007; Barklem et al. 2005; Ivans et al. 2006; Christlieb et
al. 2008; Hayek et al. 2009; Roederer et al. 2010), the model
parameters can be obtained. The results of the [Fe/H], [Eu/Fe],
[Sr/Fe], [Sr/Eu], the component coefficients and $\chi^2$ are listed
in Table 1.

Figure 2 shows our calculated best-fit results. In order to
facilitate the comparisons of the predicted abundances with the
observations, the element observed abundances are marked by full
circles. For most stars, it can be seen that in consideration of
the observational errors, there is good agreement between the
predictions and the data for all elements starting with O to Pb in
sample stars. In the top panel of Figure 3, we show individual
relative offsets ($\triangle log\varepsilon$) for the sample stars
with respect to the predictions from the parametric model. Typical
observational uncertainties in $log\varepsilon$ are $\sim0.2-0.3$
dex (dotted lines). It is clear from the Figure 2 and 3 that the
elemental abundances are in agreement for all sample stars, and
they follow closely the calculated curve. The root mean square
offset of these elements in $log\varepsilon$ shown in bottom panel
is mostly smaller than 0.20 dex for the comparison with the
predictions. This value is consistent with zero, given the
combined uncertainties in stellar abundances and predicted
abundances, confirming the validity of the parametric model
adopted in this work.

It is interesting to notice from these Figure 2 and 3 that the model
predictions are as well for the lighter neutron-capture elements as
heavier neutron-capture elements, especially for Sr, Y, and Zr. This
means that the weak r-process abundance pattern for $38\leq Z
\leq47$ adopted in this work is remarkably stable from star-to-star,
though the overall level of enrichment with respect to iron (e.g.,
[Eu/Fe]) shows a very large star to star scatter. The very large
scatter of [Eu/Fe] from star-to-star implies that very metal-poor
halo stars sample a largely unmixed early Galaxy. Since these stars
are thought to exhibit the abundance pattern produced by a single or
at most a few r-process events in the early Galaxy, the stability of
the observed abundance pattern and the good agreement with the
calculated results imply that not only main r-process events
generate a stable abundance distribution but also weak r-process
events create another universal abundance distribution.

\subsection{Example: HD 221170}

As an example, it is interesting to investigate a possible
explanation of the parameters obtained for a r-rich star HD 221170
using parametric model. Main r-process mainly occurs in the SNe
explosion of $\sim8-11 M_{\odot}$ stars (Qian \& Wasserburg 2007,
Wanajo, \& Ishimaru 2006) and the weak r-process may occur in the Fe
core-collapse SNe explosion of $\sim12-25 M_{\odot}$ massive stars
(Izutani et al. 2009). So these two processes can produce a
significant fraction of the heavy element abundances in early times
of the Galaxy. We notice from Table 1 that the component
coefficients of weak r- and main r-process for HD 221170 are
$C_w^{'} =3.755$, $C_m^{'} =0.612$, respectively, which implies that
this star is not ``pure" main r-process star or ``pure" weak
r-process star. For most of the sample stars, we can obtain the
similar result as that of HD 221170, except five stars, which are
main r-process stars (i.e., CS 22892-052, CS 29497-004 and CS
31082-001) or weak r-process stars (i.e., HD 122563 and HD 88609).

It was possible to isolate the contributions corresponding to the
weak r- and main r-process by our parametric model. Taking the
values of $C_w^{'}$, and $C_m^{'}$ into Eq. (1), the abundances of
all the neutron-capture elements in HD 221170 are obtained as listed
in Table 2. Columns (3), (4), and (5) of Table 2 give the total
abundances, the weak r-process component, and the main r-process
component in this halo star, respectively. In columns (6) and (7),
we list the weak r- and main r-process fractions (or associated with
the weak r- and main r-process) to the total abundances, these
values are useful in understanding the relative contributions of two
processes for a given neutron-capture or light element in HD 221170.
The Sr and Eu abundances are most useful for unravelling the sites
and nuclear parameters associated with the weak r- and main
r-process correspond to those in extremely metal-poor stars
(Francois et al. 2007), polluted by material with a few times of
nucleosynthetic processing. An interesting question is the behavior
of lighter r-process elements below Ba. In the weak r-process star
HD 122563 based on our model, the elemental abundances of Sr, and Eu
(the representative for pure main r-process elements) consist of
significantly different combinations of weak r- and main r- process
contributions, with r-weak:r-main ratios for Sr and Eu of 93:7 and
0:1, respectively. We explored the contributions of weak r- and main
r-process for Sr in HD 221170. Clearly, for the star listed in Table
2, the r-weak:r-main ratio for Sr is 47:53, which is smaller than
the ratio in HD 122563. The Sr abundances of HD 221170 is a result
of pollution from the two r-processes material of the former SNe. It
is suggested that even though the main r-process is dominantly
responsible for synthesis of the lighter neutron-capture elements
for this star, the contributions from the weak r-process to lighter
neutron-capture elements are not negligible at all, such as Sr, Y
and Zr. For this sample star, the heavier neutron-capture elements
are predominated, within the observational errors, by main r-process
because the r-weak:r-main ratios are mostly smaller than 10:90. The
agreement of the model results with the observations for heavier
neutron-capture elements indicates that the heavier elements are
produced by the main r-process that produces a universal abundance
pattern with fixed element ratios consistent with the solar
r-process abundance pattern. This is a similar conclusion that has
been drawn from the abundance pattern of the few very metal-poor,
strongly main r-process-enhanced stars where a large range of
elemental abundance has been determined (see, e.g., the reviews by
Truran et al. 2002; Cowan et al. 2006). It is obvious from Table 2
that the lighter r-process elements behave very differently to
heavier elements. Clearly, for the star listed in Table 2, the
r-weak:r-main ratios from Sr through Ag are larger than 20:80, which
are larger than the ratios of the heavier r-process elements.
Because the abundances of lighter neutron-capture elements are
interpreted by the mixture of weak r- and main r-process, the
agreement of the model results with the observations for the lighter
elements implies that the weak r- process creates another universal
and unique abundance pattern, which is consistent with the
conclusion obtained by Montes et al. (2007).

It is important to simultaneously analyse the observed light
elements and neutron-capture elements to investigate the physical
origins that could reproduce the abundance pattern of all observed
elements in HD 221170. The associated r-weak:r-main ratios for the
light elements from O to Zn are mostly larger than 80:20. This
implies that the production of the weak r-process elements appears
to be associated with that of Fe group elements and other light
elements (e.g., Na, Mg, Al, Si, Ca, Sc, and Ti). By combining the
analysis of neutron-capture elements, we find that the abundance
pattern of light elements for HD 221170 is close to the pattern of
the weak r-process star (e.g., HD 122563), and heavier r-process
elements is very similar to main r-process star (e.g., CS
31082-001), while the lighter neutron-capture elements can be
fitted by mixing of weak r- and main r-process material. The
element abundance pattern of HD 221170 can be explained by a star
formed in a molecular cloud that had been polluted by both weak
r-process and main r-process material. This implies that this star
should be a ``weak r + main r star".

\subsection{ Solar-system abundances and reduced component coefficients}

Generally, the solar system abundances distribution is regarded as
a standard pattern and the heavy element abundances of Population
I stars are often expected to have the similar distribution,
namely, $N_i=N_{i\odot}\times Z/Z_{\odot}$, where Z is the
metallicity of stars. The neutron-capture element abundances in
Solar-system material are sums of the two neutron-capture
processes, knowing the s-process contributions allows the
determination of the r-process contributions, or residuals, for
individual isotopes (K$\ddot{a}$ppeler, Beer \& Wisshak 1989,
Arlandini et al. 1999, Cowan \& Sneden 2006). By comparing the
r-process elemental abundances in solar system material with the
elemental abundances in metal-poor stars, we can determine the
relative contributions from the weak r- and main r-process
components to the synthesis of the neutron-capture elements in
solar system and investigate some issues to neutron-capture
nucleosynthesis at different metallicity. Therefore, we adopted
equation (1) to fit solar system r-process abundances
\begin{equation}
\begin{array}{rcl}
N_{i,r,\odot}=C_{w,\odot}^{'}N_{i,\ rw}+C_{m,\odot}^{'}N_{i,\ rm}.
\end{array}
\end{equation}
The main r-process pattern was normalized to the heavier
neutron-capture elements from Eu to Pb in solar system abundance.
And $C_{w,\odot}^{'}$ can be obtained by the lighter
neutron-capture pattern. In Figure 4, the calculated results are
compared with the solar system r-process abundances. For most
elements, the calculations produced by the equation are agreement
well with the r-process abundances of solar system (Arlandini et
al. 1999). Thus, we can approximately adopt $C_{w,\odot}^{'}N_{i,\
rw}$ and $C_{m,\odot}^{'}N_{i,\ rm}$ to represent weak r-process
component and main r-process component in solar system,
respectively.

Based on our calculation, in the Sun, the elemental abundances of
Sr, Y and Zr consist of similar combinations of weak r- and main
r-process contributions, with r-week:r-main ratios for Sr, Y and Zr
of 0.390:0.610, 0.464:0.536 and 0.462:0.538, respectively. The
average ratio of Sr, Y and Zr is 0.788. The yield of Eu per main
r-process event is estimated to be about $3\times10^{-7}M_{\odot}$
(Ishimaru et al. 2004). The yield of Sr per SN event due to the main
r-process can be estimated to be $4.5\times10^{-6}M_{\odot}$, using
the ratio of Sr/Eu ($\approx25$, Sneden et al. 2003) in the main
r-process star CS22892-052.

Clearly, the solar inventory is the result of contributions from
many stellar sources over Galactic history. Due to mixing of
nucleosynthetic products from different sources in the ISM over
this long history, the patterns for the gross abundances of a wide
range of elements in other stars with approximately solar
metallicity are observed to be close to the solar pattern. It is
well recognized that this apparent ``universality" does not imply
the existence of a single process for making all the nuclei in
nature. However, based on our calculation, the overall solar
r-pattern can be considered to have resulted from a mixture of
these two types of r-process event.

The sites of weak r- and main r-process usually discussed was an
explosive environment such as some region inside a core-collapse
SN (Qian \& wasserburg et al. 2007). Whether the elements of the
Fe group and those of intermediate mass above N can be produced by
a core-collapse SN is closely related to the pre-SN structure of
the progenitor. There are two possible venues leading to a
core-collapse SN: (1) collapse of an Fe core produced by
progenitors of 12-25$M_{\odot}$, (2) collapse of an O-Ne-Mg core
produced by progenitors of 8-11$M_{\odot}$. Observational data on
low-metallicity stars in the Galactic halo show that sites
producing the heavier neutron-capture nuclei do not produce Fe or
any other elements between O and Zn. Insofar as a forming
core-collapse SN is key to producing the heavy r-nuclei, then the
mostly possible sources are SNe resulting from collapse of O-Ne-Mg
cores, because they do not produce the elements of the Fe group or
those of intermediate mass (above C and N). This suggests that the
main r-process elements cannot be produced by massive stars of
$>11M_{\odot}$, which result in Fe core-collapse SNe and are
sources for the elements from O to Zn (Woosley et al. 2002). On
the other hand, the weak r-nuclei with A$\sim$90-110 are in
general present in metal-poor stars with low abundances of heavier
neutron-capture nuclei and produced in conjunction with the
elements from O to Fe group elements (Qian \& Wasserburg 2002,
2007; Izutani et al. 2009). Thus, this implies that the origins of
weak r-process are Fe core-collapse SNe from progenitors of
12-25$M_{\odot}$, ejected most of their nucleosynthetic products,
such as a host of nuclei from O to the Fe group(Qian \& Wasserburg
2007). This has been shown to be possible in the calculated
results of Izutani et al.(2009), which can reproduce the
observational data of Sr, Y, and Zr in addition to the elements
from O to Zn in weak r-process stars.

Using our calculated results, we discuss more thoroughly the
mixing of weak r- and main r-process in the solar system. Assuming
a Miller et al.(1979) initial mass function (IMF), the ratio of an
Fe core-collapse SN event from a progenitor of 12-25$M_{\odot}$ to
an O-Ne-Mg core-collapse SN event from a progenitor of
8-12$M_{\odot}$ is 0.797, which is approximately equal to
calculated r-weak:r-main ratio for Sr, Y and Zr in the solar
system. This means that a similar amount of lighter
neutron-capture nuclei is produced per event for weak r- and main
r-process venues. Thus, the yield of Sr per weak r-process event
is estimated to be about $4.5\times10^{-6}M_{\odot}$, which is in
good agreement with theoretical estimates from explosive
nucleosynthesis calculations of weak r-process elements in
extremely metal-poor core-collapse SN (Izutani et al. 2009).
Obviously, in our picture, the ratio has a straightforward
explanation in terms of the IMF. With the same IMF, we would find
frequency ratios of 1.156, when taking 11 $M_{\odot}$ instead of
12$M_{\odot}$as the threshold between the two classes. Obviously,
this estimate requires more stringent constraints on the masses
from theoretical models and more accurate observed statistics.

It is interesting to define reduced component coefficients of
metal-poor stars, $C_w$ and $C_m$ as:
\begin{equation}
N_{i}=(C_{w}N_{i,\ rw,\odot}+C_{m}N_{i,\ rm,\odot})10^{[Fe/H]},
\end{equation}
where $N_{i,\ rw,\odot}=C_{w,\odot}^{'}N_{i,\ rw}$ and $N_{i,\
rm,\odot}=C_{m,\odot}^{'}N_{i,\ rm}$. In equation (4), we choose the
solar component coefficients as a standard and assume all of them
are equal to 1. Note that the reduced component coefficient of
metal-poor star gives the relative contribution of the individual
process to that of solar system, normalized to the value of [Fe/H].
In fact, if we substitute $10^{[Fe/H]}$ with $Z/Z_{\odot}$ and
assume $C_w=C_m=1$, equation (4) will be returned to the r-process
component of $N_{i,r}=N_{i,r,\odot}\times Z/Z_{r,\odot}$, which have
used by Aoki et al. (2001).

\subsection{Discussion of reduced Component Coefficients}

The reduced component coefficients calculated are listed in the
column (8) and (9) of Table 1. We can obtain important information
about the neutron-capture nucleosynthesis from the reduced
component coefficients, summed up as follows :

1. The component coefficients $C_w$ and $C_m$ represent the relative
contributions from the weak r-process and the main r-process to the
heavy element abundances, respectively. With them, we can accurately
determine the relative contributions of the individual r-process to
the neutron-capture element abundances in metal-poor stars and then
compare them with the corresponding component coefficients of the
solar system in which $C_w=C_m=1$. Thus, $C_i>1$ (i=r or m) means
that, except for the effect of metallicity, the contributions from
the corresponding r-process to the neutron-capture element
abundances in metal-poor stars is larger than that in the solar
system. $C_i<1$ means that, except for the effect of the
metallicity, the contributions from the corresponding r-process to
the neutron-capture element abundances in metal-poor stars is less
than that in the solar system. If $C_w$ and $C_m$ are not equal to
each other, then the relative contribution from the weak r-process,
the main r-process to the neutron-capture element abundances are not
in the solar proportions.

2. From Table 1 we note that, for most sample stars except weak
r-process stars and main r-process stars, both $C_w$ and $C_m$ are
larger than unity. This means that the contributions from the weak
r- and main r-process to the neutron-capture element abundances in
these stars is larger than that in the solar system. Thus, similar
to HD 221170, these stars can be called ``weak r + main r stars".

3. By comparing the values of $C_w$ and $C_m$, we can select those
stars with special neutron-capture element abundance distributions
and study them individually. If one component coefficient of one
metal-poor star is much larger than another and larger than unity,
this star might have been formed in a Galactic region that were not
well mixed chemically and the corresponding neutron-capture process
may be dominantly responsible for the neutron-capture elements in
this star. In this case, we should consider only the contribution of
this neutron-capture process to the abundances of neutron-capture
elements in this star. For example, the ultra metal-poor halo star
CS 22892-052 is a star of this situation. With our model, the
component coefficients in this star are $C_w$=0.0, $C_m$=49.204,
which mean that the main r-process is responsible for the synthesis
of neutron-capture elements in the star. In Table 1 we can find that
main r-process stars (i.e., CS 22892-052, CS 29497-004 and CS
31082-001) and weak r-process stars(i.e., HD 122563 and HD 88609)
can be attributed to two different extreme categories. It should be
noted that our model will be not effective for stars with higher
metallicities ([Fe/H]$\gtrsim$$-1.6$), because the contribution from
s-process cannot be ignore at this metallicity (Travaglio et al.
1999). The component coefficients $C_w$ and $C_m$ with [Eu/Fe], as
illustrated in Figure 5, contain some interesting information for
the enrichment of r-process in sample stars. Clearly, this figure
reveals different groups that can be distinguished by the component
coefficients.

Our model is based on the observed abundances of two main
r-process stars and two weak r-process stars, and the observed
abundances in metal-poor stars, so all the uncertainties of those
observations will be involved in the model calculations, which
results in the errors of the calculations and the uncertainties of
the model. The errors of the calculations simply result from the
measurement errors in those sample stars. These measurement errors
are only reflected in equation (2) and can be estimated by the
model. According to measurement errors of the abundance ratios in
two main r-process stars and two weak r-process stars, we estimate
that the typical errors of the model calculations are about
$\pm0.2$ dex, because the errors of the model calculations are
approximately equal to the measurement errors. Considering the
uncertainties of the model calculations, one can find that there
are more elements whose abundances are explained by the model
predictions. However, we find that all these uncertainties cannot
explain the larger errors of elements, such as Lu in BD
+17$^{\circ}$3248 and Ag in HD 221170. This implies that the
understanding of the true patterns of the weak r-process and main
r-process is incomplete for at least some of these elements.

\section{ Conclusions}

The chemical abundances of the metal-poor star are an excellent test
bed to set new constraints on models of neutron-capture processes at
low metallicity. It did not appear that a straightforward ab initio
solution to the r-process problem was within reach. On the other
hand, a abundance-analysis approach based on the available
observational data might offer some helpful guidance. Stars with low
[Fe/H] values would provide important clues to whether and how
core-collapse SNe are associated with the r-process. Based on the
observation of metal-poor stars and neutron-capture element
nucleosynthesis theory, we set up a model to determine the relative
contributions from weak r- and main r-process to the neutron-capture
element abundances in metal-poor stars. With this model we calculate
the  elemental abundances in 14 sample stars and the component
coefficients $C_w$ and $C_m$. Considering the Eu overabundance and
excesses of lighter neutron-capture elements in most sample stars,
it is worth to compare abundances of these stars with those of weak
r-process stars and main r-process stars. Such study can provide
clues for the understanding of the enrichment in neutron-capture
elements of metal-poor stars, given that they should conserve the
characteristics of r-process nucleosynthesis in massive stars in the
early galaxy. The neutron-capture and light elements abundance
pattern of most sample stars could be explained by a star formed in
a molecular cloud that had ever been polluted by weak r- and main
r-process material. Overall, the predicted abundances fit well the
observed abundance patterns of these stars, from the light
elements(O through Zn) and lighter neutron-capture elements (Sr
through Ag), as well as heavier neutron-capture elements (Ba through
Pb). We find that the main r-process source is responsible for the
heavier neutron-capture elements such as Eu; and the light elements,
such as Fe group, accompany with the weak r-process production,
while both sources produce the lighter neutron-capture elements. The
main r-process elements nearly are not produced in conjunction with
all light elements from O to Fe group elements, which is similar to
the observed results of main r-process stars. The abundance pattern
of light elements for the stars is very close to the pattern of weak
r-process stars. This implies that these stars should be ``weak r +
main r star". The abundance, derived from the our model, are in
excellent agreement with the observed values of sample stars, which
means that the uniform and unique abundance pattern of weak
r-process have extended to [Eu/Fe]$\approx1.0$, [Fe/H]$\approx-2.1$,
and all observed elements.

In closing, we remind the reader that the results here have the
virtue of being model independent, in that they do not refer to
detailed r-process nucleosynthesis calculations. Of course, a full
solution of the weak r-process will demand such calculations. Our
hope is that the results here will provide a useful guide in
interpreting those more complete r-process models. Our results give
a constraint on the models of the r-process that yielded lighter and
heavier neutron-capture elements in the early Galaxy. We look
forward to a large data set on weak r- and main r-process abundances
in very low metallicity stars, which will improve our understanding
of weak r- and main r-process nucleosynthesis in the early Galaxy.

\section*{Acknowledgments}

This work has been Supported by the National Natural Science
Foundation of China under Nos 10673002, 10973006, 10847119 and
10778616, the Natural Science Foundation of Hebei Provincial
Education Department under Grant No 2008127£¬ the Science Foundation
of Hebei Normal University under Grant No L2007B07 and the Natural
Science Foundation of Hebei Proince under Grant no. A2009000251.

\bsp

\label{lastpage}

\end{document}